\documentclass[11pt,letterpaper, twocolumn]{article}
\usepackage{authblk}
\usepackage[version=3]{mhchem} 
\usepackage[T1]{fontenc}       
\usepackage{graphicx}
\usepackage{dcolumn}
\usepackage{bm}
\usepackage{enumitem,lipsum}
\usepackage{chemarr}
\usepackage[mathlines]{lineno}
\usepackage[export]{adjustbox}
\usepackage{float}
\usepackage[caption=false]{subfig}
\usepackage{rotating} 
\usepackage[svgnames]{xcolor}	
\usepackage{wasysym}
\usepackage{textcomp}
\usepackage{anysize}
\marginsize{2cm}{2cm}{1cm}{2cm}

\title{Generalized Fick-Jacobs Approach for Describing Adsorption-Desorption Kinetics in  Irregular Pores under Non-Equilibrium Conditions}
\author{Aldo Ledesma-Dur\'an}
\author{Sa\'ul Iv\'an Hern\'andez-Hern\'andez}
\author{Iv\'an Santamar\'ia-Holek}
\affil{Unidad Multidiscliplinaria de Docencia e Investigaci\'on-Juriquilla, Facultad de Ciencias, Universidad Nacional Aut\'onoma de M\'exico, CP 76230, Juriquilla, Quer\'etaro, Mexico
}
\date{}

\begin{document}

\maketitle



\begin{abstract}
 We present a study exploring the range of applicability of a generalized Fick-Jacobs equation in the case when diffusive mass transport of a fluid along a pore includes chemical reactions in the bulk and pore's surface. The study contemplates non-equilibrium boundary conditions and makes emphasis on the comparison between the predictions coming from the projected Fick-Jacobs description and the corresponding predictions of the original two-dimensional mass balance equation, establishing a simple quantitative criterion of validity of the projected description. For the adsorption-desorption process, we demonstrate that the length and the  local curvature of the pore are the relevant geometric quantities for its description, allowing for giving very precise predictions of the mass concentration along the pore. Some schematic cases involving adsorption and chemical reaction are used to quantify with  detail the concentration profiles in transient and stationary states involving equilibrium and non-equilibrium situations. Our approach provides novel and important insights in the study of diffusion and adsorption in confined geometries.
\end{abstract}


\section{Introduction\label{sec:intro}}

The transport of mass in confined geometries poses interesting open questions that need to be addressed in order to improve our understanding of many processes taking place in biology, chemistry, nanotechnology and biomedicine \cite{koch1990diffusion,Arico2005,pannada2006,adiga2008,yusko2011controlling,Hoffman2012,ledesma2014multiscale}. 
Pertinent examples are the transport of molecules in channels of biological membranes
\cite{zwanzig1992diffusion,reguera2001kinetic,berezhkovskii2007diffusion,burada2007biased,dagdug2010unbiased,martens2011entropic,garcia2015covariant}, as well as in technological systems ranging from zeolites, carbon nanotubes and serpentine channels in microfluidic devices, to artificially produced pores in thin solid films  \cite{millhauser1988diffusion,karger1992diffusion,hidenori1995,wijmans1995solution,yan2008,martinez2015onsager}. 

Since the movement of molecules through a pore is strongly confined in every dimension except one, which is termed as the longitudinal, in general terms the mass transport inside this system can be considered as a quasi-one dimensional process.  A successful one-dimensional mathematical description of these systems may enormously simplify many technical and computational questions allowing,  for further generalizations, the incorporation of interactions among the transported particles as well as the presence of adsorption-desorption \cite{santamaria2012non,santamaria2013entropic} and chemical reaction kinetics \cite{biktasheva2015drift,martens2015front} . In addition, it may lead to improve the formulation of models for the effective transport properties of the materials in these strongly confined situations \cite{kalinay2005projection,bradley2009diffusion,martens2011entropic,pineda2014projection,dorfman2014assessing}. 

 Similar considerations lead to many recent studies to formulate projected descriptions  in which an effective one-dimensional Fokker-Planck equation for the single particle probability density of finding a particle along the direction of the pore  $p(x,t)$, is obtained. This leads to the  Fick-Jacobs equation \cite{zwanzig1992diffusion}
\begin{equation}\label{eq:fjoriginal}
\frac{\partial p}{\partial t}=\frac{\partial}{\partial x}\left[ D_x\,\omega\frac{\partial }{\partial x}\left(\frac{p}{\omega}\right)\right].
\end{equation}
 The integration of the original Fokker-Planck equation along the transverse coordinates allows to keep the influence of the pore walls on the motion of the particles through the incorporation of the imposed boundary conditions in terms of the position dependent parameter $\omega(x)$, which accounts for the width of the pore, see Fig. \ref{figure1}, and the space dependent effective diffusion coefficient $D_x=D(x)$, that accounts for the reduction to just one coordinate of the mean square displacement of the original molecular diffusion coefficient $D_0$ \cite{reguera2001kinetic,kalinay2005projection,bradley2009diffusion,martens2011entropic,pineda2014projection}.  
 
A number of hypothesis behind these projection procedures are explicitly and tacitly assumed. The main supposition allowing the obtention of  a closed form for the projected equation is that the diffusion is slow enough, in such a way that the characteristic relaxation time for the transport along the pore is much grater than the time of relaxation in the transverse direction. A second crucial hypothesis is that the pore walls are impermeable to diffusion. With these hypothesis, the so called Fick-Jacobs equation is obtained  for the  probability density $p(x,t)$  \cite{zwanzig1992diffusion}.

The Fick-Jacobs equation (\ref{eq:fjoriginal}) has been studied in several particular cases of pore geometry, giving satisfactory results when compared with Brownian motion simulations and predicted diffusion coefficients \cite{berezhkovskii2007diffusion,martens2011entropic,pineda2014projection,burada2009diffusion,dagdug2010unbiased}. The advantage of using a reduced equation is not only the fact that it is easier to solve than the two dimensional diffusion or Fokker-Planck-Smoluchowski equations, but mainly, that provides information on the relevant geometric elements for the transport in these tube-structures. One important example, is the relation of the effective diffusion coefficients with the specific shape of the pore\cite{zwanzig1992diffusion,reguera2001kinetic,kalinay2005projection,bradley2009diffusion,pineda2014projection}. 
Another interesting topic where the Fick-Jacobs scheme has been used is when diffusion takes place in presence of chemical reactions inside the pore\cite{martens2015front,biktasheva2015drift}. These recent models try to reproduce some experimental and simulation results obtained for reaction diffusion systems in confined media, which are reported in the literature\cite{agladze2001electrochemical,ginn2004microfluidic,kitahata2004slowing,pagliara2014channel}.

In view of the previous considerations, in this work we derive a generalization of the projected one-dimensional transport equation of the Fick-Jacobs type that can be used to describe the important case in which isothermal chemical reactions  \cite{martens2015front,biktasheva2015drift} as well as adsorption-desorption processes in the pore walls may take place \cite{santamaria2012non,santamaria2013entropic}.
In short, we will follow a similar projection methodology as typically used to obtain the Fick-Jacobs equation. Nevertheless, we will average the mass balance equation that includes general chemical reaction kinetics. Hence, our description is more appropriately represented in terms of the concentration of particles rather than in terms of the probability density. The advantage of this approach is that it is applicable to any concentrated fluid that obeys a Fick law of diffusion of the form
\begin{equation}\label{eq:diffusion}
\frac{\partial C}{\partial t}=\nabla\cdot\left[D_x\nabla C(x,y,t)\right],
\end{equation}
where $C$ represents a concentration of particles in  mol/cm$^2$. 

In the following sections, we will show that the generalized Fick-Jacobs equation has the form \cite{santamaria2012non,santamaria2013entropic} 
\begin{equation}\label{eq:fjivan}
\frac{\partial c}{\partial t}=\frac{\partial}{\partial x}\left[ D_x\,\omega\frac{\partial }{\partial x}\left(\frac{c}{\omega}\right)\right]+j(x,t),
\end{equation}
where now $c(x,t)$ is a linear concentration of particles in mol/cm, and  $j(x,t)$ is a generalized flux that depends upon all the possible fluxes of particles inside the pore (other than the diffusive), including chemical reactions. The two main contributions of this work are to demonstrate that the chemical flux $j(x,t)$ is not uniform along the pore\cite{santamaria2012non,santamaria2013entropic}, but it is related to the curvature of the pore walls, and to show the importance of this fact in the description of adsorption-desorption processes.

Finally, we have tested the obtained generalized Fick-Jacobs equation by comparing the numerical solutions of the last equation and the original two-dimensional mass balance equation under non-equilibrium  conditions. This will allow us to show the importance of the pore curvature in the determination  of the regions inside the pore where the material is adsorbed or desorbed more rapidly. 

The article is organized as follows. In Section \ref{sec:deduction}, we summarize the general aspects of the mass balance equation in a pore including the different boundary conditions of the problem.
In Section 3 we deduce a generalized Fick-Jacobs equation that permit us to quantify the bulk and surface mass distribution of a fluid with reactant species along a pore where diffusion, adsorption-desorption and chemical reactions may take place.  In Section \ref{sec:numerico}, we obtain numerical solutions of the generalized Fick-Jacobs equation and compare them with the direct solution of the two dimensional mass balance equation. This permit us to establish the limits of applicability of the reduction scheme in terms of quantities such as the length-width ratio, tortuousness and constriction of the pore. Section 5 is devoted to exemplify the  performance of the generalized Fick-Jacobs equation in situations involving mainly adsorption and chemical reactions under non-equilibrium conditions. The main goal of this section is to prove the importance of the pore curvature in the description of adsorption-desorption processes, and its relation with the localization of active sites along the pore. In Section \ref{sec:conclusion} we discuss our results, as well as the general scope and advantages of our model. Finally, we sketch some possible extensions for future work.


\section{Mass balance in pores\label{sec:deduction}}
\begin{figure}[btp]
\centering
\includegraphics[width=8cm]{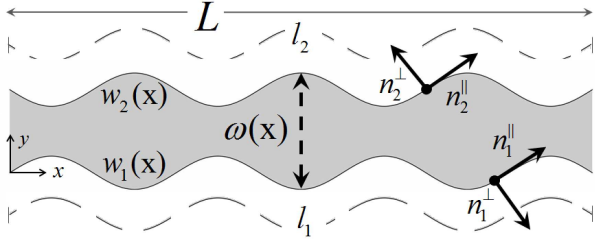}
\caption{Geometry of a pore of length $L$. The width of the pore $\omega(x)$ is defined through  the functions $w_1(x)$ and $w_2(x)$, that represent the pore walls, in Eq. (\ref{eq:defomega}). The unit parallel and perpendicular vectors to the surface are represented by $\hat{n}^{\parallel}$ and $\hat{n}^{\perp}$ , respectively. The arc lengths of each wall are $l_1$ and $l_2$.  The particular geometry  depicted was obtained by using the values of $(L,\tau,\kappa)=(5,0,0.2)$ (all values in cm) in Eq. (\ref{eq:porefunctions}). The width of pore end is $d=1$cm. This parameter, together with the molecular diffusion coefficient  $D_0=d^2/$1s fixed the spatio-temporal scale for subsequent cases, unless otherwise indicated.
 \label{figure1}}
\end{figure}
Let's consider a gas, whose molecules diffuse inside a pore whose geometric characteristics are shown in Fig. \ref{figure1}. The interior of the pore, in gray, is delimited by the boundaries $w_1(x)$ and $w_2(x)$ which are continuous and differentiable functions of the longitudinal coordinate $x$ in such a way that  the pore's width $\omega(x)$ is given by 
\begin{equation}\label{eq:defomega}
\omega(x)=w_2(x)-w_1(x).
\end{equation}  
If  the molecules can be adsorbed and desorbed at the pore walls, then the total flux of molecules in the pore, $\mathbf{J}(x,y,t)$ having dimensions of mol/cm$\cdot$s, is given as the sum of the contributions
\begin{equation}\label{eq:totalflux}
\mathbf{J}(x,y,t)=\mathbf{J}^{diff}+\mathbf{J}^{surf},
\end{equation}
\noindent where $\mathbf{J}^{diff}$ is the diffusive flux and $\mathbf{J}^{surf}$ is the surface flux of material due to chemical interactions with the pore's walls. 

The diffusive flux is defined by
\begin{equation}\label{eq:fluxdiff}
\mathbf{J}^{diff}(x,y,t)=-D\nabla C(x,y,t),
\end{equation}
where $C(x,y,t)$, in mol/cm$^2$, represents the concentration of the fluid inside the pore, and $D$ is the so-called Fickean-diffusion coefficient \cite{karger1992diffusion}. 
Furthermore, we will assume the well known boundary conditions for this flux near the walls of the pore, namely, that the diffusive flux is parallel to the walls  \cite{zwanzig1992diffusion,kalinay2005projection,bradley2009diffusion,pineda2014projection}
\begin{equation}\label{eq:cfdiff}
\mathbf{J}^{diff}(x,w_i,t)=A_i(x,y)\hat{\mathbf{n}}^{\parallel}_{i}, 
\end{equation}
\noindent where $\hat{\mathbf{n}}^{\parallel}_{i}(x,w_i)$ is the unit vector parallel to the surface $w_i$,  see Fig. \ref{figure1}, and $A_i(x,y)$ is an unknown function (in mol/cm$\cdot$s) for each one of the two surfaces, hereafter numbered with indexes $i=1,2$. 

The transport of molecules through the pore walls, $\mathbf{J}^{surf}(x,y,t)$, is produced by adsorption-desorption processes and, therefore, it vanishes in every internal point of the domain, that is,
\begin{equation}\label{eq:nonreact}
\mathbf{J}^{surf}(x,y,t)=0 \hspace{1cm} w_1(x)<y< w_2(x),
\end{equation}
whereas, when evaluated at the boundaries, it takes the form
\begin{equation}\label{eq:cfchem}
\mathbf{J}^{surf}(x,w_i,t)=r_{i}(x,t)\hat{\mathbf{n}}^{\perp}_i(x,w_i),  \hspace{1cm} i=1,2,
\end{equation}
where $r_i$, in mol/cm$\cdot$s, is the kinetic rate of input or output of material at each wall and 
$\hat{\mathbf{n}}^{\perp}_i(x,w_i)$ is the unit vector normal to each  pore surface, see Fig. \ref{figure1}.  In this context, we refer as surface flux as the one from the bulk to the surface or vice versa depending if it is a process of adsorption or desorption. It does not have to be confused with the surface diffusive flux that we are not considering in this model and, unlike the adsorptive flux, is parallel to the surface. We have assumed that the surface flux $\mathbf{J}^{surf}$ is perpendicular to the pore's wall. Notice that the subscript $i=1,2$ indicates that each surface could have different production rates. This means that we are including cases in which the adsorption-desorption processes do not occur necessarily at the same rate or in the same sites on the two walls. We will illustrate this in Section 5.

The general mass balance equation inside the pore can be expressed as
\begin{equation}\label{eq:fick2}
\frac{\partial C}{\partial t}+\nabla\cdot \mathbf{J}=G(x,y,t).
\end{equation}
where the chemical production or consumption inside the bulk of the fluid is given by $ G(x,y,t)$, having dimensions of mol/s$\cdot$cm$^2$. 

The novelty of this scheme is  the explicit separation of the chemical flux at the surface of the walls $\mathbf{J}^{surf}$, from the chemical production inside the pore given by $G$ and occurring at the bulk of the fluid. It will be shown later that this approach distinguishes between different functional dependencies for both terms.

Substituting  Eq. (\ref{eq:fluxdiff}) in the total flux defined by Eq. (\ref{eq:totalflux}) and then the result into Eq. (\ref{eq:fick2}), one obtains the general mass balance equation
\begin{equation}\label{eq:fickgral}
\frac{\partial }{\partial t}C(x,y,t)=\nabla\cdot[D_x\nabla C]-\nabla \cdot \mathbf{J}^{surf}+G.
\end{equation}
This differential equation should be complemented with the appropriate boundary conditions. At the pore's walls the diffusive flow $\mathbf{J}^{diff}$ should satisfy Eq. (\ref{eq:cfdiff}), whereas for the surface flux $\mathbf{J}^{surf}$, the appropriate boundary conditions are given through Eq. (\ref{eq:cfchem}). 

The boundary conditions at the pore ends depend upon the particular conditions to which the porous material is subjected. In Section \ref{sec:numerico}, we will exemplify three important cases, namely  a) the pore is closed at both ends:
\begin{equation}\label{eq:membrana}
\frac{\partial }{\partial x}C(0,y,t)=\frac{\partial }{\partial x}C(L,y,t)=0, 
\end{equation}
b) the porous material is immersed in a fluid that enters from both ends until the pore is filled
\begin{equation}\label{eq:membrana2}
C(0,y,t)=C_0=C(L,y,t),
\end{equation}
and c) the porous material acts as a membrane that separates regions with different concentrations:
\begin{equation}\label{eq:membrana3}
C(0,y,t)=C_0, \hspace{.2cm}C(L,y,t)=C_L.
\end{equation}
In all cases $x=0$ and $x=L$ determine the pore ends, see Fig. \ref{figure1}. 




\section{ Projection of the mass balance equation: generalized Fick-Jacobs equation and geometrical aspects }

The projection of the mass balance equation can be done on the assumption that the transport along the longitudinal coordinate $x$ takes a larger time than the characteristic time of transport along the transverse direction. Depending upon the specific pore geometry (constriction, tortuosity, periodicity), the validity of the projection Fick-Jacobs scheme has been previously quantified through the use of convergence parameters of power series. These parameters are expressed in terms of the ratio of two quantities and some examples of them are: the ratio between the longitudinal and transverse components of the diffusion coefficient \cite{kalinay2005projection,pineda2014projection}; the ratio between the channel averaged width and its period \cite{laachi2007force,yariv2007electrophoretic}, the ratio between the corrugation degree and the period in a sinusoidal pore \cite{martens2011entropic} or, the square ratio between characteristic longitudinal/transversal length \cite{bradley2009diffusion}. As our purpose in this work is not centered in this convergence, we should use a more restrictive version of the fourth criterion already mentioned: $w(x)/L<<1$,  see Fig. \ref{figure1}. As we will see in the Section devoted to numerical simulations, this criterion is enough for the pores and initial and external conditions that we have studied in this work.

Following the usual procedure, we integrate Eq. (\ref{eq:fick2}) with respect to the coordinate $y$ from $w_1(x)$ to $w_2(x)$. The result is 
\begin{align}\label{eq:smolu8}
\frac{\partial c }{\partial t} =\frac{\partial }{\partial x}\left\{D_x\left[\frac{\partial c}{\partial x} -  C(w_2)w'_2+C(w_1)w'_1\right]\right\} \nonumber \\
\hspace{2.5cm} +j_s(x,t)+g(x,t),
\end{align}
where we have used the notation $w'_i= dw_i/dx$. The first term at the right hand side of this exact equation is the reduced diffusion contribution in which the reduced concentration $c$, having units of mol/cm, has been defined by
\begin{equation}\label{eq:candC}
c(x,t)=\int_{w_1(x)}^{w_2(x)}C(x,y,t)dy.
\end{equation}
The third and fourth terms of Eq. (\ref{eq:smolu8}) are the contributions coming from the boundary terms after using Leibniz integration rule.  The reduced chemical production $g(x,t)$ is given by
\begin{equation}\label{eq:redg}
g(x,t)=\int_{w_1(x)}^{w_2(x)}G(x,y,t)dy,
\end{equation}
\noindent and finally, the reduced surface flux $j_s(x,t)$, in mol/s$\cdot$cm, takes the form
\begin{equation}\label{eq:redj}
j_s=\left[J_y(w_2)-J_x(w_2)w'_2\right]-\left[J_y(w_1)-J_x(w_1)w'_1\right],
\end{equation}
where the subindexes $x,\,y$ stand for the first and second coordinates of the flux defined in Eq. (\ref{eq:totalflux}). It should be noticed that Eq. (\ref{eq:smolu8}) contains the original concentration $C$ as well as the reduced concentration $c$.  Usually, in order to put Eq. (\ref{eq:smolu8}) in terms of only the reduced concentrations $c$, a couple of considerations have to be pointed out. 

\subsection{Rapid transverse relaxation}
The first assumption is based on the consideration that the pore is long enough. In this case, the changes on the concentration take a longer time along the longitudinal coordinate than in the transverse one, and therefore we may assume $C(x,y,t)\approx C(x,t)$, see Refs. \cite{zwanzig1992diffusion,reguera2001kinetic}. Hence, from Eq. (\ref{eq:candC}), we have 
\begin{equation}\label{eq:approx}
c(x,t)= C(x,t)\omega(x).
\end{equation}
Using this approximation, the first term in the equation Eq. (\ref{eq:smolu8}) can be simplified to get a similar expression to that of Eq. (\ref{eq:fjivan}), this is,
\begin{equation}\label{eq:smolu14}
\frac{\partial c}{\partial t}=\frac{\partial}{\partial x}\left[ D_x\,\omega\frac{\partial }{\partial x}\left(\frac{c}{\omega}\right)\right]+j_s(x,t)+g(x,t).
\end{equation}
In the dilute limit, with absence of any chemical interaction, the Eq. (\ref{eq:smolu14}) reduces to the Fick-Jacobs equation, Eq. (\ref{eq:fjoriginal}) \cite{zwanzig1992diffusion,reguera2001kinetic}. Now, if there are some kind of chemical processes, we recover expression (\ref{eq:fjivan}) where $j$ represents indistinctly both $j_s$ and/or $g$ \cite{santamaria2012non,santamaria2013entropic}.  This generalized Fick-Jacobs equation, as well as the diffusion equation Eq. (\ref{eq:diffusion}) are solely valid in the absence of external forces, solvent flow \cite{martens2013hydrodynamically,martens2014giant}, as well as complex electrochemical potentials \cite{malgaretti2014entropic,malgaretti2016entropically}.

\begin{figure*}[btp]
\centering
\includegraphics[width=16cm]{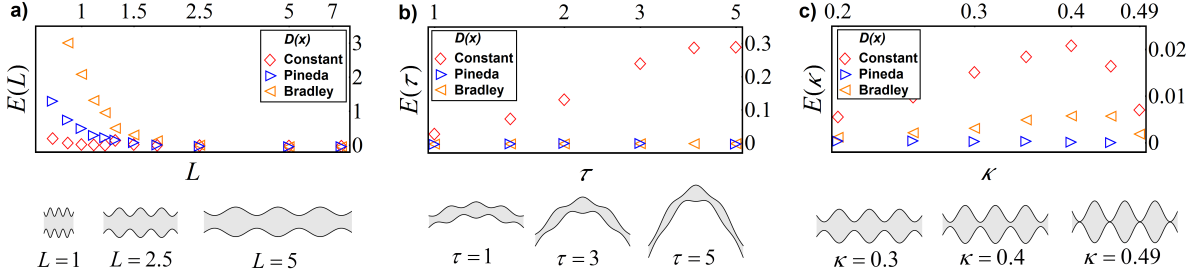}
\caption{ Error measured for several pores according to (\ref{eq:error}). The plots correspond to the variation of the length of the pore $L$, its tortuousness $\tau$, or its constriction $\kappa$, respectively. In each plot, three different effective diffusion coefficients $D_x$ in the Fick-Jacobs equation are used (\ref{eq:fjoriginal}): Constant coefficient ($\diamond$), Pineda coefficient \cite{pineda2014projection} ($\triangleright$) and Bradley coefficient \cite{bradley2009diffusion}($\triangleleft$). In this  figure, $N=100$ and $(L,\tau,\kappa)$ are the same as in Fig. \ref{figure1}, except the  particular parameter that is taken as variable. The value of $t_r$ is the fifth part of the stationary time $t_s$, and it is different for each point in the graph.
\label{figure2}}
\end{figure*}

In this work, we will consider as a first approximation that the transport properties of the particles on the bulk, measured by $D_x$, are mainly determined by the existence of hard walls. This means that the effective diffusion coefficient is mainly determined by this confinement and, therefore, that the existence of chemical reactions does not affect it considerably. This allows us to use previous expressions reported in literature for this effective coefficient\cite{bradley2009diffusion,pineda2014projection}.  However, in the general case, this assumption is not necessarily true. The presence of reactions and adsorption processes could affect the dynamics of the particles inside the pore (see, for example, Refs. \cite{shapiro1987chemically,brenner2013macrotransport}). This modification of the effective diffusion coefficient has to be considered in systems where the (bulk or surface) reactions produce faster changes in the local concentration than those produced by diffusive transport. In other words, this means that the use of a projected scheme, as the one presented in this work, is restricted to diffusion and reaction parameters that accomplish the condition that the characteristic time of diffusion along the $x$ direction ($\tau_x$) is larger than the characteristic time of reaction ($\tau_r$). As the former time is also larger than the characteristic time of transport along $y$, ($\tau_y$), our model is restricted to the accomplishment of this hierarchy.

\subsection{Reduced surface flux and generalized Fick-Jacobs equation}
The second consideration is related to the fact that the terms contributing to the reduced surface flux $j_s$ and appearing in Eq. (\ref{eq:redj}), contain the components of the total flux defined in Eq. (\ref{eq:totalflux}) but evaluated at the walls of the pore. Thus, it is important to prove that the parallel component of the total flux does not contributes to the reduced surface flux defined in  (\ref{eq:redj}), that is, only the perpendicular component of the total flux contributes to $j_s$.

In view of the boundary conditions, Eqs. (\ref{eq:cfdiff}) and (\ref{eq:cfchem}), we may first split  the total flux at the wall in its parallel and perpendicular components, that is,
\begin{equation}\label{eq:Jtrabajado}
\mathbf{J}(x,w_i,t)=(\mathbf{J}\cdot \hat{n}^{\parallel}_i )\hat{n}^{\parallel}_i+(\mathbf{J}\cdot \hat{n}^{\perp}_i )\hat{n}^{\perp}_i,
\end{equation} 
where we have introduced the external parallel and perpendicular unit vectors at the walls of the pore, whose mathematical expressions are given by 
\begin{equation}\label{eq:nparallel}
\hat{n}^{\parallel}_i=\frac{1}{\gamma_i}(1,w'_i),\,\,\,\,\,\text{and}\,\,\,\,\,\hat{n}^{\perp}_i=\frac{(-1)^i}{\gamma_i}(-w'_i,1),
\end{equation}
for $i=1,2$, see Fig. \ref{figure1}. These expressions naturally introduce the parameter $\gamma_i$ which is defined by 
\begin{equation}
\gamma_i(x)=\sqrt{1+(w'_i)^2(x)},
\end{equation}
and represents an arc-length density of the pore's wall measuring its curvature and length, since it satisfies the condition
\begin{equation}\label{eq:arc}
\int_0^L \gamma_i(x)dx=l_i,
\end{equation}
where $l_i$ is the arc length of each wall.

Using now the boundary conditions, Eqs. (\ref{eq:cfdiff}) and (\ref{eq:cfchem}) into (\ref{eq:Jtrabajado}), we obtain 
\begin{equation}\label{eq:Jtrabajado2}
\mathbf{J}(x,w_i,t)=A_i\hat{n}^{\parallel}_i+r_a\hat{n}^{\perp}_i,
\end{equation}
which, after using in turn the expressions given in Eqs. (\ref{eq:nparallel}) for the parallel and normal vectors, leads finally to the concise expression 
\begin{equation}\label{eq:redj2}
j_s(x,t)=r_1(x,t)\gamma_1(x) +r_2(x,t)\gamma_2(x) .
\end{equation} 
It follows from this equation that the reduced surface flux is characterized solely by the perpendicular component of the total diffusion flow at the walls of the pore and, more important, it is weighted by the length density $\gamma_i$. In real systems, the total rate associated to adsorption-desorption reactions clearly depends on the density of active sites at the pore's wall. In this sense, Eq. (\ref{eq:redj2}) appears to be an excellent way to theoretically model and quantify this fact, since the addition of the products $r_i(x,t)\gamma_i(x)$ suggests that the total adsorption-desorption reaction is proportional to  the parameter $\gamma_i$. Modeling appropriately the factor $\gamma_i$ will allow to describe very accurately surface catalyzed chemical reactions. We will discuss this point in detail in Section 5.

Substitution of  Eq. (\ref{eq:redj2}) into Eq. (\ref{eq:smolu14}) yields to
\begin{equation}\label{eq:fjparac}
\frac{\partial c}{\partial t}=\frac{\partial}{\partial x}\left[ D_x\,\omega\frac{\partial }{\partial x}\left(\frac{c}{\omega}\right)\right]+\sum_i^2 r_i \gamma_i(x)+g,
\end{equation}
which is a generalized Fick-Jacobs  equation that considers the existence of bulk and surface catalyzed chemical reactions. Eq. (\ref{eq:fjparac}) condensates the most relevant aspects of the longitudinal mass transport in a pore, including diffusion, adsorption and chemical reactions. 
We should remark that, in the derivation of Eq. (\ref{eq:fjparac}), we have included in the chemical terms $r_i$ and $g$ the functional dependency on the spatial coordinate $x$ and the time. This also implies that they can be given in terms of the reduced concentration itself $c(x,t)$, the arc function $\gamma(x,t)$ or some other geometrical factor of these variables. In fact, as $r_i(x,t)$ measures the rate of adsorption/desorption at the wall, which usually depends upon the concentration near the walls, the explicit dependence on the concentration has the form $r_i(x,t) \to r_i(C(x,w_i,t),x,t)$. In similar lines,  $g$ could represent fixed sources and sinks inside the pore, as well as a chemical reaction which depends upon the concentration itself. In the last case, the explicit dependence can be written as $g(x,t) \to g(c(x, t),x,t)$. In order to abbreviate our notation, we will write only the dependence in $x,y$ and $t$, but keeping in mind that we are considering both cases.

Although this projection of the 2D mass balance equation to a reduced 1D equation could  represent some loss of information, the advantages of the generalized Fick-Jacobs equation, Eq. (\ref{eq:fjparac}), are remarkable. First, this reduced equation can be solved numerically in a simpler way than the two-dimensional mass balance equation and, second, the projection procedure explicitly reveals the relevant geometric aspects for the transport along those materials, information that can be useful in the determination of other thermodynamic quantities, such as effective rates of reaction or effective diffusion coefficients in irregular confined systems \cite{santamaria2013entropic,pineda2014projection}.   

In the following sections, we will test the application range of Eq.  (\ref{eq:fjparac}) by directly comparing its predictions (numerical solutions) with those  arising from the two-dimensional mass balance equation Eq. (\ref{eq:fickgral}) for different physical conditions.


\section{Validity conditions of the Fick-Jacobs equation. Diffusion \label{sec:numerico} }

In this section we will establish a quantitative criterion that permit us to know the range of applicability of the generalized Fick-Jacobs equation, Eq. (\ref{eq:fjparac}), in the case when chemical reactions are absent. This analysis is necessary because in the deduction of Eq. (\ref{eq:fjparac}) all the approximations were performed in order to obtain the Fick-Jacobs diffusion operator. Thus, as a consequence of our analysis, we can be confident that the predictions of the generalized Fick-Jacobs equation are reliable or, at least, they do not fail due to the inadequacy of the hypothesis underlying the projection procedure for some geometries, or the particular effective diffusion coefficient used.
\begin{figure*}[btp]
\centering
\includegraphics[width=16cm]{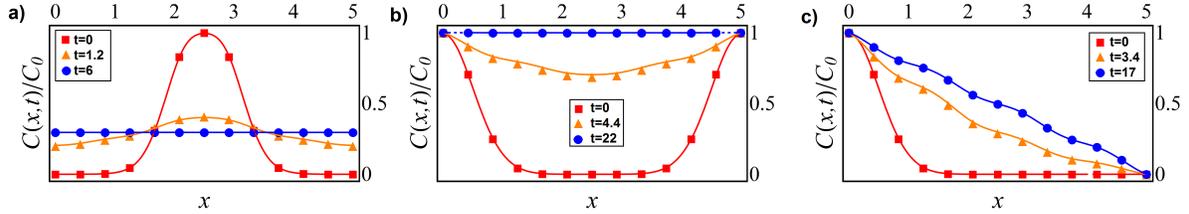}
\caption{ Comparison between single dimensional mass concentrations $c_{2D}(x,t)$ (symbols) and $c_{FJ}(x,t)$ (lines) arising from the numerical solutions of Eqs.  (\ref{eq:diffusion})  and (\ref{eq:fjoriginal}), respectively. For both equations, the diffusion coefficient was taken as constant, $D_x=D_0=1$cm$^2$/s.  The geometry is the same as that described in Fig. \ref{figure1}. The three figures represent the three boundary conditions stated in Eqs. (\ref{eq:membrana})-(\ref{eq:membrana3}) respectively, using $C_0=1$ and $C_L=0$. In each subplot, the concentration is represented at three different times: the initial condition $t_0$ (filled squares), the time at which the stationary profile is reached $t_s$ (filled circles), and a transient time $t_r=t_s/5$ (filled upward triangles). This same indication is valid for all subsequent examples. In these cases, $t_s=6,22$ and $17$s, respectively.\label{figure3}}
\end{figure*}

The quantitative criterion is formulated in terms of the error committed by the numerical solution $c_{FJ}(x,t)$, of the projected equation Eq. (\ref{eq:fjparac}), with respect to the projection, $c_{2D}(x,t)=\int_{w_1}^{w_2} C_{2D}(x,y,t) dy$, of the complete numerical solution $C_{2D}(x,y,t)$ of the  two-dimensional mass balance equation, Eq.  (\ref{eq:diffusion}). Mathematically, this error can be measured as the quotient between the  residual sum of squares and the total sum of squares as \cite{spiegel1975schaum} 
\begin{equation}\label{eq:error}
E(L, \tau, \kappa, t_{r}, D_x)=\sum\limits_{j=1}^{N}\frac{[c_{2D}(x_j,t_{r})-c_{FJ}(x_j,t_{r},D_x)]^2}{[c_{2D}(x_j,t_{r})- \langle c_{2D}(t_{r}) \rangle ]^2}.
\end{equation}
This expression measures the difference in the solutions over $N$ points distributed homogeneously along the pore. The comparison is made at an intermediate time $t_r$ that we have chosen as a fifth part of the stationary time, due to the fact that near the stationary time, the profiles scarcely change and the comparison does not reflect the variation of the reduced scheme for transient times, which are of interest as well. Finally, $\langle c_{2D}(t_{r}) \rangle$ is the averaged two-dimensional reduced solution in all the $N$ points. This quantity provides a scale between the difference of solutions and the characteristic value of the concentration for each pore. In this way, an error value near zero corresponds to a very good approximation of the reduced solution, whereas a value near 1 corresponds to an error of the same magnitude that the characteristic concentration. The value of the error $E(L, \tau, \kappa, t_{r}, D_x)$ provides a global criterion that permits  to discern when the reduced scheme is applicable or not, especially  if one is interested in short transient times. Furthermore, for a given pore geometry,  this criterion permit us to choose the suitability of the different expressions of the effective diffusion coefficient that have been reported in the literature.  We will test only the diffusion coefficients reported in Refs. \cite{bradley2009diffusion,pineda2014projection} because they incorporate information about the transversal section area of the pore as well as of the specific form of the center line of the channel. The diffusion coefficients obtained in more general studies \cite{ogawa2013diffusion,valdes2014fick}  considering three dimensional pores have to be validated in two-dimensional geometries in order be of utility for a study like ours.

In order to solve the diffusion equation in two dimensions, Eq. (\ref{eq:diffusion}), we have used the standard Finite Element Method (FEM) for space discretization and a BDF solver for the time integration. The mesh used in each case was chosen extremely fine, with about 6000 elements, in order to guarantee a mesh average quality larger than 0.95. The iterative regression starts with an initial time step of $T_{max}/100$ and a number of iterative steps guaranteeing a relative error tolerance of 0.0001. In other hand, the numerical solution of the Fick-Jacobs equation, Eq. (\ref{eq:fjparac}) was obtained using an adaptative mesh generator for the spatial coordinate and an implicit time integrator. Inasmuch as no analytical solution of the model equations exists for the problems studied here, we are assuming that the difference between both of solutions [estimated with Eq. (\ref{eq:error})] is much larger than the error of the numerical solution of each model with respect to the exact solution. The same methods were used to obtain all the results reported. 
 It  is convenient to emphasize that the error proposed in Eq. (\ref{eq:error}) is a difference between the numerical solutions of Eqs. (\ref{eq:fickgral}) and (\ref{eq:fjparac}) for several pore geometries. Since this error is zero in the equilibrium state and we are interested on intermediate times, we choose an intermediate time to measure this difference, $t_r = t_{s}/5$ with $t_{s}$ the time elapsed to reach a stationary profile. The magnitude of $t_s$ depends on several factors, like the geometry of the channel and the external conditions. The dependence on the parameters $L,\tau$, and $\kappa$ is implicit in the solution of the concentration profile $c_{FJ}$, that in turn depends on the particular form of the effective diffusion coefficient used.

Fig. \ref{figure2} shows the  estimated errors (open symbols) for a sinusoidal shaped pore reflecting three different physical situations of relevance that are controlled by means of the pore's length $L$, tortuousness $\tau$, and constriction $\kappa$.  These errors were calculated for different effective diffusion coefficients: Constant coefficient (diamonds, $\diamond$), Pineda coefficient \cite{pineda2014projection} (right triangles,  $\triangleright$), and Bradley coefficient \cite{bradley2009diffusion} (left triangles,$\triangleleft$). The basic geometry of the studied pore is defined by the relations (see, Fig.  1)

\begin{subequations}\label{eq:porefunctions}
\begin{equation}
w_1(x)= \tau \sin \frac{\pi x}{L}-\kappa \sin \frac{7\pi x}{L},
\end{equation}
\text{and}
\begin{equation}
w_2(x)=d+\tau \sin \frac{\pi x}{L}+ \kappa\sin \frac{7\pi x}{L}.
\end{equation}
\end{subequations}
In these expressions, the parameter $d$ regulates the size of the pore at the ends and it was fixed at the value of 1cm. This parameter, together with the molecular diffusion coefficient $D_0=d^2/s=1$cm$^2/$s, provides all the space-temporal characteristic scales for comparing to other lengths and times.

With the aim to simplify the analysis of the error of the reduced scheme as a function of the geometrical parameters $\tau$ and $\kappa$, we first search for a pore in which the Fick-Jacobs model can be used regardless the effective diffusion coefficient, that is, a pore in which the length $L$ is long enough (compared with $d$) in such a way that the concentration profiles obtained by numerical solution of Eq. (\ref{eq:fjparac}) are not significantly different when a constant coefficient or an  effective diffusion (which depends upon $x$) is used, see Fig. \ref{figure3}. This allowed us to assume a constant diffusion coefficient (the same as in a non-confined situation). For the pore defined by Eqs. (\ref{eq:porefunctions}) and illustrated in Fig. \ref{figure1}, this was achieved with the parameters $(L,\tau,\kappa)=(5,0,0.2)$. The comparison between the one-dimensional concentration profiles $c_{2D}(x,t)$ (closed symbols) and $c_{FJ}(x,t)$ (lines)  shown in Fig. \ref{figure3} is very good, even for transient times, in all the boundary and initial conditions considered [see Eqs. (\ref{eq:membrana})-(\ref{eq:membrana3})].

In Fig. \ref{figure2}.a) we have fixed all parameters as in Fig. 3, except the length $L$. We have calculated the error $E$ for different lengths using Eq. (\ref{eq:error}). From the results it follows that the Fick-Jacobs equation can be used with less than 5\% of error in a sinusoidal pore with any of the diffusion coefficients considered for pores obeying the condition: $L/d>2$. We can remark that both effective diffusion coefficients provide good results even when they were deduced assuming equilibrium conditions \cite{kalinay2005projection}.

In Figs.  \ref{figure2}.b) and \ref{figure2}.c) we show the error obtained when changing the tortuousness $\tau$ and its constriction $\kappa$, respectively. These results illustrate an important question regarding the spatial dependence of the diffusion coefficient. For instance, Bradley coefficient was derived by incorporating in an appropriate form the effects of tortuousness of a pore but not the net effects of constriction. As a consequence, as shown in Fig. \ref{figure2}.b), it leads to the same predictions as Pineda's coefficient, but fails in the case when constriction is increased, see Fig. \ref{figure2}.c).

The main conclusion of this section is the following: The Fick-Jacobs equation can be used in a wide variety of pore geometries with sinusoidal shape, even for non-equilibrium initial and boundary conditions, if the pore obeys the aspect ratio condition already mentioned.  In addition, for different tortuousness and constrictions, the predictions of the concentration profiles by the Fick-Jacobs equation are be in good approximation with the predictions of the two-dimensional mass balance equation, as far as one uses the adequate effective diffusion coefficient. We want to emphasize  that we have proven that the predictions of the concentration profiles by the  Fick-Jacobs equation are valid not only for stationary or equilibrium situations, but even for short transient times.

\section{Applications of the generalized Fick-Jacobs equation: Adsorption-desorption kinetics}

In this section, we will discuss two important aspects related with the application of the generalized Fick-Jacobs equation to adsorption-desorption kinetics in a pore. First, we will discuss the pertinent variables of the theoretical description that allow the connection  with the experiments. The second aspect is the importance of the geometrical parameter $\gamma$ that regulates the intensity of the adsorption-desorption process and can be related with the distribution of active sites in the pore's surface. This last point will be illustrated with a series of examples based on limiting conditions of a Langmuir adsorption-desorption kinetics.

\subsection{Average concentration. Average production}\label{sec:average}

In the context of physical and chemical adsorption, it is more convenient to 
write Eq. (\ref{eq:fjparac}) in terms of averaged quantities, because the adsorption-desorption rates inside the pore are known in terms of interactions per second and unit of volume. In view of this, we define the averaged concentration
\begin{equation}\label{eq:avC}
\mathcal{C}(x,t)=\frac{1}{\omega}\int_{w_1}^{w_2} C(x,y,t)dy,
\end{equation}
the averaged chemical production  
\begin{equation}\label{eq:avG}
\mathcal{G}(x,t)=\frac{1}{\omega}\int_{w_1}^{w_2} G(x,y,t)dy,
\end{equation}
and the averaged surface-reaction rate 
\begin{equation}\label{eq:randR}
\mathcal{R}_i(x,t)=\frac{r_i(x,t)}{\omega(x)},
\end{equation}
with $i=1,2$ for each surface. 

If the condition of long pore is accomplished, then we can use again Eq. (\ref{eq:approx}) in order to rewrite Eq.  (\ref{eq:fjparac}) in terms of  the average quantities $\mathcal{C}$, $\mathcal{G}$ and $\mathcal{R}_i$. The resulting equation is
\begin{equation}\label{eq:fjgralC}
 \frac{\partial \mathcal{C}}{\partial t}(x,t)=\frac{1}{\omega}\frac{\partial}{\partial x}\left[ D_x\,\omega \frac{\partial \mathcal{C}}{\partial x} \right]+\sum_i^2\gamma_i(x) \mathcal{R}_i+\mathcal{G}.
\end{equation}

The last equation describes the same physical situation as Eq. (\ref{eq:fjparac}) and therefore, is also an approximation of the original system obeying Eq. (\ref{eq:fickgral}). 
 A similar form of this equation has been used in Refs. \cite{martens2015front,biktasheva2015drift} in the case when  adsorption and desorption processes are absent, i.e., when $r_i=0$. However, in the context of adsorption in porous media, the importance of Eq. (\ref{eq:fjgralC}) relies in the fact that the statement of the problem and its interpretation reflects more faithful the experimental measurement capabilities. This is because the initial and boundary conditions of the partial differential equation are more easily stated in terms of the original concentration, as illustrated in Eqs. (\ref{eq:membrana})-(\ref{eq:membrana3}). In addition, the chemical rates and the corresponding chemical kinetics of the adsorption-desorption processes inside a porous media are usually known as average quantities over the pore and, therefore, are functions of $\mathcal{C}$. Beside this, the surface concentrations are usually determined through volume quantities. This  fact allows one to use $\mathcal{R}_i$ for volumetric concentrations instead of the quantity $r_i$, which in turn is used for surface concentrations [see Eq. (\ref{eq:cfchem})]. 

Let us illustrate this situation using the classical example of the Langmuir adsorption-desorption process\cite{ruthven1984principles,carberry1987chemical}. 
For this particular example, the average adsorption-desorption rate along the pore is given by the well known formula \cite{fogler1999elements,cussler2009diffusion}: 
\begin{equation}\label{eq:langmuir}
R=k_{ads} C_{bulk} (C_{max}-C_{surf} )-k_{des}C_{surf},
\end{equation}
where $C_{bulk}$ represents the mass concentration of the particles at the bulk, and $C_{surf}$ the mass concentration of the particles already adsorbed at the wall. The corresponding rate constants of adsorption and desorption are represented by $k_{ads}$ and $k_{des}$, respectively. Finally, $C_{max}$ represents the maximum mass concentration that the pore can adsorb on its surface. It should be emphasized that this description does not contains any specific information about the width of the pore or the width of the adsorbed layer, since $C_{max}$ and $C_{surf}$ are determined over the entire porous material, and therefore do not represent surface concentrations (in units of mol per unit of length in our case). 
\begin{figure*}[]
\centering
\includegraphics[width=14cm]{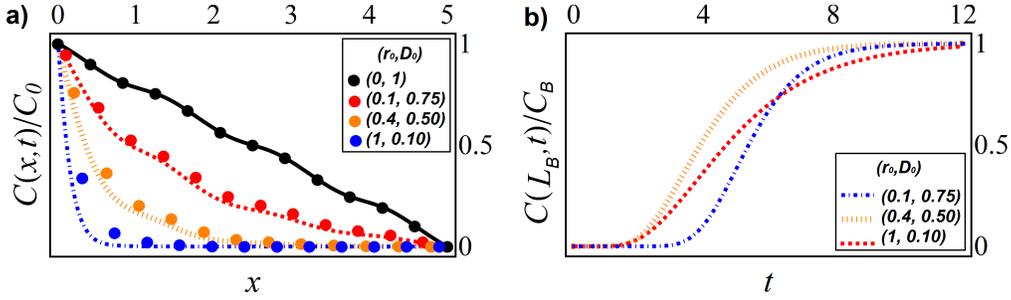}
\caption{ Characterization of the adsorption kinetics given in Eq. (\ref{eq:radsor}) for the probe pore defined by Eqs. (\ref{eq:porefunctions}).  The geometry is the same as that described in Fig. \ref{figure1}. a) Normalized numerical stationary solutions of Eq. (\ref{eq:fjgralC}) with boundary conditions given by Eq. (\ref{eq:membrana3}) with $C_0=1$ and $C_L=0$. We assumed different values for the reaction velocity constant and the difusion coefficient: $(k_{ads},D_0)$, in units of $1/$s for $k_{ads}$, and cm$^2$/s for $D_0$. The filled symbols ({\tiny$\newmoon$}) correspond to the numerical solution of the two-dimensional mass balance equation, Eq. (\ref{eq:fickgral}). b) Breakthrough curves as a function of time for a site located at a distance of  $L_B = 9/10 L$ from the  pore inlet. The maximum of concentration at this point is $C_B = C(9/10 L,t\to\infty)$. \label{figure4}}
\end{figure*}  

For pores satisfying the condition $L/d>2$, our previous results allow us to use a reduced scheme such as that on Eq. (\ref{eq:fjgralC}) and, therefore, to change the total rates and concentrations for they averaged values along the vertical coordinate. Thus, we may replace the capital letters ($C$, $R$ and $G$) by calligraphic ones ($\mathcal{C}$, $\mathcal{R}$ and $\mathcal{G}$) and use the relations Eqs. (\ref{eq:approx}) and (\ref{eq:randR}), in order to rewrite Eq. (\ref{eq:langmuir}) in terms of reduced quantities. The resulting expression is
\begin{equation}\label{eq:langmuir2}
r \simeq k_{ads} \frac{c_{bulk}}{\omega} (c_{max}-c_{surf} )-k_{des}c_{surf}.
\end{equation}
Unlike the term $R$, appearing in Eq. (\ref{eq:langmuir}) and representing the averaged reaction rate along the pore, the reduced reaction rate $r$ represents the chemical reaction rate exactly at the surface of the pore, see Eq. (\ref{eq:cfchem}). As a consequence of this, Eq.(\ref{eq:langmuir2}) represents a Langmuir-like scheme for the reduced concentrations in which the adsorption rate becomes weighted by the factor $1/\omega$. A similar result has been obtained previously in Ref. \cite{santamaria2013entropic} on the basis of irreversible thermodynamics applied directly to the reduced description. In that case, the form of the chemical rate was deduced directly by assuming that the chemical potential of the particles at the bulk  $\mu(x,t)_{bulk}$, contains an entropic barrier correction of the form  $\Delta S = k_B \ln{{\omega}/{\omega_0}}$. The corresponding resulting expression was
\begin{equation}\label{mu}
\mu_{bulk}=k_B T \ln{\frac{c_{bulk}}{c_0}}-k_B T\ln{\frac{\omega}{\omega_0}},
\end{equation}
where $T$ is the temperature, and $c_0$ and $\omega_0$ are the reference reduced mass concentration and width, respectively, corresponding to the case of a regular pore. Eq. (\ref{mu}) is compatible with our derivation and it implies that the projection of the system to a single dimension necessarily entails the addition of correction terms in the form of entropic barriers which are originated in the boundaries of the system \cite{zwanzig1992diffusion,santamaria2013entropic}. 

In the context of our model, the explanation of the factor $1/\omega$ in the adsorptive term of Eq. (\ref{eq:langmuir2}) lies directly in the geometric aspects of the reactive interaction between the particles at the bulk and the surface. To explain this, let us consider any point $x_0$ on the wall of the pore, and a slim slice of width $\Delta x$. Then, the probability of adsorption in the site $x_0$ is proportional to the number of particles near the surface at that point. In the reduced description, this probability is proportional to the probability of finding a particle in the slice of area $\omega(x_0)\Delta x$. In the diluted case, it could be assumed that this probability is inversely proportional to this area and, therefore, inversely proportional to $\omega(x_0)$. This reflects the obvious fact that the probability of one adsorption event in one point along $x$ is, in the reduced description, inversely proportional to the pore width at that point.

Summarizing, the distinction between reduced and average quantities is very important since the chemical rate productions at the surface (quantified by $r_i$) could be unknown. Notwithstanding, the average rates of production quantified by $\mathcal{R}_i$ could be measured by means of more simple uptake experiments \cite{deutschmann2009heterogeneous,roque2012adsorption,karger2012diffusion}.
One of the merits of the present study is that it establishes the exact mathematical relations between these two descriptions and, as a consequence of this, it establishes a theoretical bridge that connects the measurements of the average quantities with the more detailed processes taking place at the surface of the pore.

\subsection{Effect of the active site density in the pore's mass distribution}
In this subsection, we present specific examples on the use of the generalized Fick-Jacobs equation in problems involving diffusion, adsorption and chemical production.

\subsubsection{Adsorption under non-equilibrium boundary conditions}
As a first example, let us consider an ideal gas inside the probe pore (see Fig. 1) in the presence of an external concentration gradient defined by the boundary conditions given in Eq. (\ref{eq:membrana3}). In a non-confined situation, the particles have a  molecular diffusion coefficient $D_0$. However, in our example, they enter to the pore and are adsorbed on the walls with a superficial rate proportional to the bulk concentration. The adsorption kinetics is therefore of the form

\begin{equation}\label{eq:radsor}
r_{ads}=k_{ads} C(x,t),
\end{equation}
where $k_{ads}$ is the adsorption velocity constant that we have assumed constant.
 
The result of the interaction between the particles and the adsorption sites is a diminution of the bulk concentration, compensated with the continuous entry of material at one end of the pore. In Figure \ref{figure4}.a), we show the stationary concentration profiles of the gas in the bulk phase for different values of $k_{ads}$ and $D_0$. As expected, the bulk concentration at each point is lower when  increasing $k_{ads}$. The local efficiency of the adsorption process is shown in the breakthrough curves of Figure \ref{figure4}.b).  The intermediate slope of the breakthrough curves increases as the quotient $k_{ads}/D_0$ is increased. This is due to the fact that the particular efficiency of adsorption in a section of the pore depends upon the balance of the chemical rate of adsorption measured by $k_{ads}$, and the local availability of particles measured by $D_0$. 
\begin{figure}[]
\centering
\includegraphics[width=8cm]{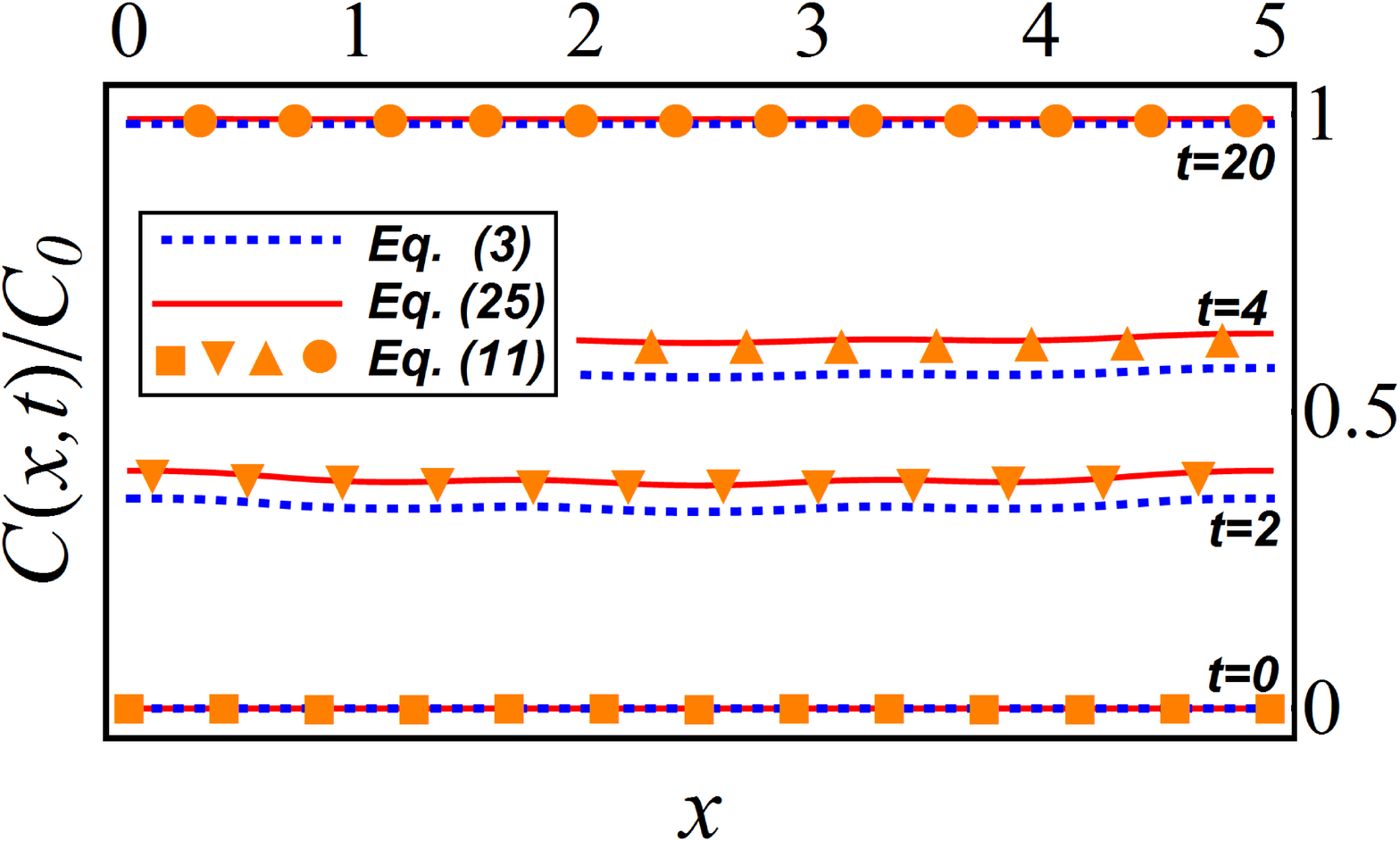}
\caption{ Time evolution of the concentration profiles along the pore for the desorption kinetics given in Eq. (\ref{des-kinet}).  The geometry is the same as that described in Fig. \ref{figure1} . The direct solution of the two dimensional mass balance equation (filled symbols) is compared with the prediction arising from the generalized Fick-Jacobs equation Eq. (\ref{eq:fjparac}) [solid lines] and that of Eq. (\ref{eq:fjivan}) from Ref. \cite{santamaria2013entropic} (dashed lines). We used the boundary conditions stated in Eq. (\ref{eq:membrana}) and the parameters $(k_{des},D_0) =(1,0.1)$. The maximum concentration inside the pore is $C_m=1$.  \label{figure5}}
\end{figure} 

Notice that Figure \ref{figure4}.a) shows an increase on the deviation of the solution obtained for the reduced scheme (lines)  with respect to the solution of the original mass balance equation (filled circles). This deviation is more noticeable as the diffusion coefficient tends to zero, due to the fact that differential problem of Eq. (\ref{eq:fjparac}) with the boundary conditions Eq. (\ref{eq:membrana3}),  is ill conditioned for $D_0 \to 0$. 

\subsubsection{Desorption kinetics and effective reaction velocity constants}
In  Fig. \ref{figure5}, we compare the evolution of the mass concentrations for the  desorption kinetics 
\begin{equation}\label{des-kinet}
r_{des}=k_{des} [C_{t}-C(x,t)],
\end{equation}
which is predicted by Eq. (\ref{eq:fickgral}) [filled symbols], Eq. (\ref{eq:fjparac}) [solid lines] and the solutions of the thermodynamic based model Eq. (\ref{eq:fjivan}) [dotted blue lines] \cite{santamaria2012non}.  We considered that, initially, all the particles are at the surface with a concentration $C_{t}$ and, for simplicity's sake, we assumed the desorption velocity constant $k_{des}$ as a constant. 

The comparison of the net change of concentration obtained by direct solution of the two dimensional mass balance equation (filled symbols), Eq. (\ref{eq:fickgral}), is very good when compared with the solution of the reduced scheme proposed in this work, Eq. (\ref{eq:fjparac}) [solid lines].  The reason is that the reduced scheme, Eq. (\ref{eq:fjparac}), introduces an effective local desorption rate 
\begin{equation}
r_{eff} (x)\equiv \gamma(x)r_{des},
\end{equation}

\noindent that introduces the pore's length as a weight taking into account the local curvatures and the real length of the walls of the pore. The comparison between a previous model, given by Eq. (\ref{eq:fjivan}) with the direct solution of the two dimensional mass balance equation is not so good because in that model the local curvature is not taken into account, and therefore:  $r_{eff}(x)=2 r_{des}$. The solution in this case tends to underestimate the magnitude of the adsorption-desorption dynamics. 


\begin{figure*}[btp]
\centering
\includegraphics[width=16cm]{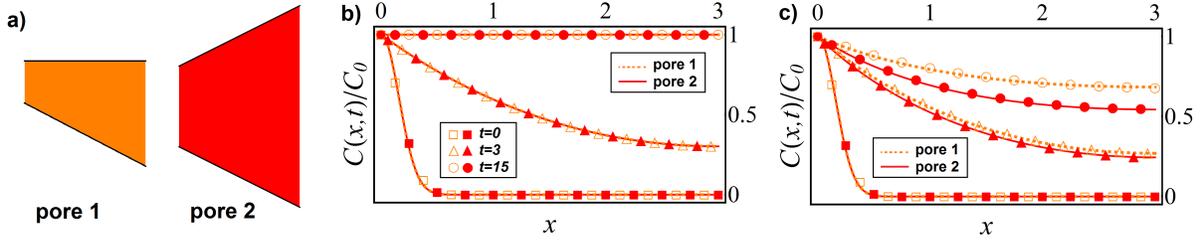}
\caption{ a) Two straight pores with the same diffusion dynamic  but different adsorption dynamics. The first one is modeled by the form functions $w_1=-Ax$ and $w_2=1$, and the second pore is given by $w_1=-Ax$ and $w_2=2+Ax$, with $A=0.5$. The length is $L=3$ for both cases. b) Identical concentration profiles in both pores for three different times (in seconds) when there are only diffusion ($D_0=1$ and $k_{ads}=0$).  c) Diffusion-adsorption profiles for ($D_0=1$ and $k_{ads}=0.1$). As $\gamma$ has   different values for each pore ($2.11$ and $2.23$ respectively), the adsorption process in the second pore  is more effective than in the first one.  \label{figure6}}
\end{figure*}

\subsubsection{Interplay between mass diffusion and the effective reaction velocity constant}
The relevance of the pore's length density $\gamma(x)$ on the adsorption kinetics can be illustrated by means of three examples. 

\paragraph{Straight boundaries of different length.}
Consider the two pores with straight boundaries shown in Figure \ref{figure6}.a). In the projected description, both pores exhibit exactly the same diffusion dynamics because they have the same effective diffusion coefficient, $D_x = 0.92\,D_0$, and obey exactly the same reduced equation. The effective diffusion coefficient was obtained by using the analytical expressions provided in Ref.  \cite{bradley2009diffusion} and \cite{pineda2014projection}. The equivalence of the diffusive processes in both pores can be also proven by comparing they numerical solutions as it is shown in Fig. \ref{figure6}.b) for three different times. In this case, for both pores, the material enters at the left side until it is filled. Both pores have the same length and are closed at the right side. 

Interestingly, when  a simple adsorption dynamics of the form defined by Eq. (\ref{eq:radsor}) is included in Eq. (\ref{eq:fjparac}), then the concentration profiles for both pores obey different adsorption kinetics since the geometric factor $\gamma$ is approximately $18\%$ larger in pore 2 than in pore 1. In Fig. \ref{figure6}.c) it is shown that the homogeneous adsorption along the pore is favored in pore 2 and, therefore, the bulk concentration profile in pore 2 (filled symbols and solid lines) is lower than in pore 1 (open symbols and dashed lines). The difference increases drastically for intermediate times and is most remarkable in the stationary profiles (circles). 

\begin{figure}[]
\centering
\includegraphics[width=8.0cm]{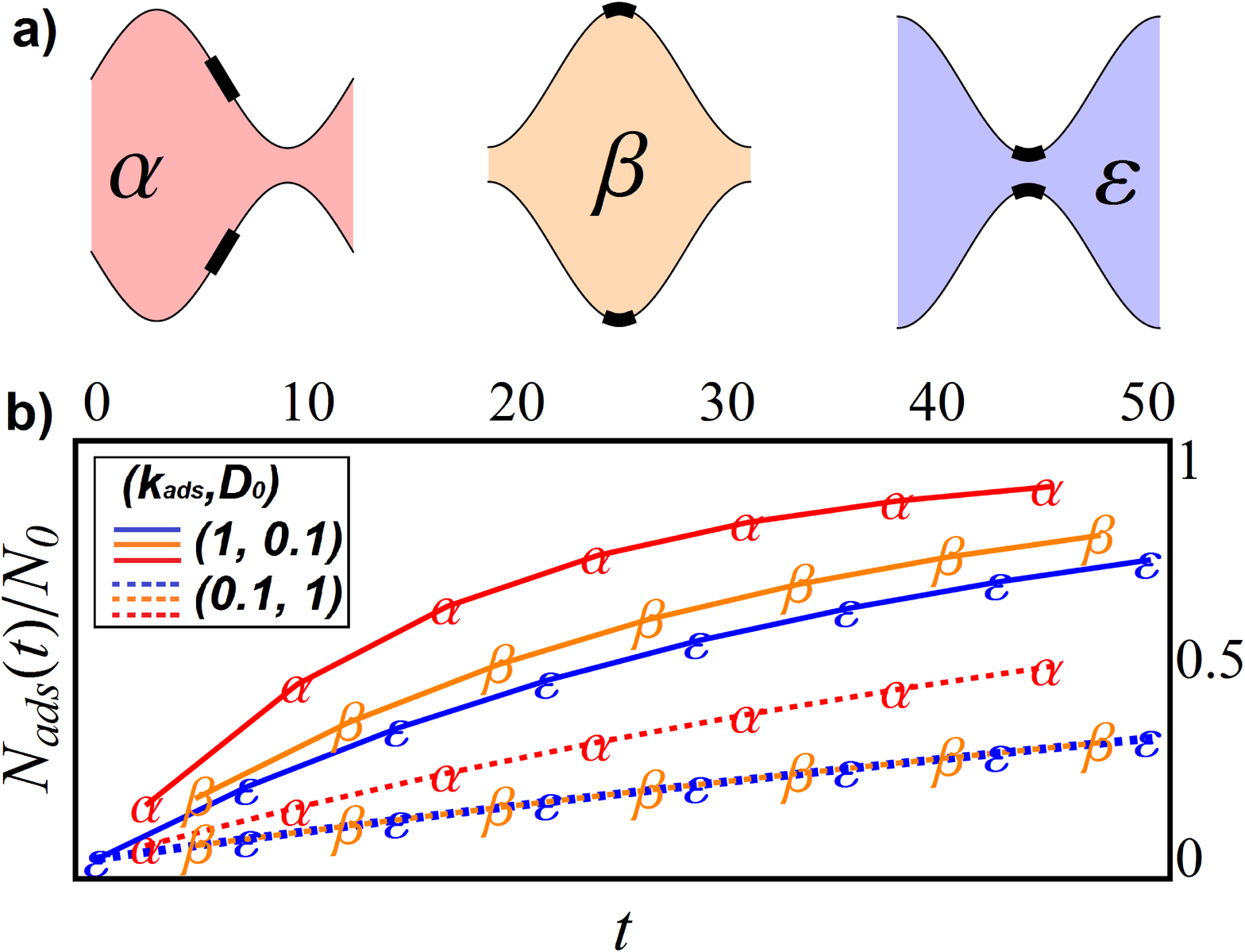}
\caption{ Adsorption dynamic in sinusoidal pores with active sites in specific places of the pore. The geometry is given by $(L,\tau,\kappa)=(5,0,0.4)$ in Eqs. (\ref{eq:porefunctions}) and the boundary conditions at the ends of the pore are taken as periodic.
a) A single period of three infinite sinusoidal pores (with periodic boundary conditions) with the same shape and different location of the active sites at the walls (emphasized thicker black lines). b) Comparison of the normalized number of adsorbed particles, $N_{ads}(t)/N_0$, as a function of time for two different diffusion-adsorption kinetics. $N_0$ is the total final number of particles, and adsorption and diffusion are characterized by the parameters 
$(k_{ads},D_0)=(1.0,0.1)$ [solid lines] and $(k_{ads},D_0)=(0.1,1.0)$ [dashed lines]. Lines correspond to solutions of the generalized Fick-Jacobs equation, whereas markers $\alpha$, $\beta$ and $\epsilon$ correspond to the solutions of the two dimensional mass balance equation.  \label{figure7}}
\end{figure}

\paragraph{Discrete distribution of active sites.} 
In order to illustrate how the specific location of the active sites affects the kinetics of adsorption, we have used an infinite sinusoidal pore. Initially, it is homogeneously filled with a gas that can be adsorbed at the walls at some discrete regions. Fig. \ref{figure7}.a) shows three possible  positions of the adsorption sites ($\alpha$, $\beta$ and $\epsilon$) emphasized with thicker black lines at the two boundaries.

This physical situation can be modeled as an adsorption reaction in which we can introduce the active length density $\gamma_{ac}(x)$.  This quantity is related with the adsorption probability in a specific site $p_{ads}(x)$, and the corresponding local curvature of the pore at this point, $\gamma(x)$. Thus, we have 
\begin{equation}\label{eq:kdesorpt2a}
\gamma_{ac}(x) = p_{ads}(x)\gamma(x).
\end{equation}
This relation is very important because it illustrates that the efficiency of the adsorption depends on the location and curvature of the active site in the pore.

For instance, in our case we have modeled the existence of localized active sites through the adsorption probability
\begin{equation}\label{eq:kdesorpt2}
p_{ads}(x) = p_0{\exp[-(x-x_k)^2/\sigma^2]},
\end{equation}
where $p_0$ is a normalization constant, $x_k$ is the central position of the active site along the pore's wall, and $\sigma$ measures its distribution over $x$ around $x_k$.  This $p_{ads}$ represents the probability of being adsorbed in a specific place measured along the $x$ coordinate.  The corresponding reaction rate entering in Eq. (\ref{eq:fjparac}) is therefore
\begin{equation}\label{ads-kinetsecVb}
r_{eff}(x)=p_0\exp[-(x-x_k)^2/\sigma^2] \,  \gamma(x) \, r_{ads},
\end{equation}
with $r_{ads}$ given by Eq. (\ref{eq:radsor}). 

Fig.  (\ref{figure7}.b) shows the number of particles adsorbed for three different positions of active sites along each period of the pore and for two different values of the ratio $k_a/D_0$. Markers correspond to the predictions of the two dimensional mass balance equation whereas the lines correspond to the generalized Fick-Jacobs approach. Solid and dashed curves correspond to two limiting behaviors in which adsorption is faster than diffusion, $(k_a,D_0)=(1.0,0.1)$, and in which diffusion is faster than adsorption, $(k_a,D_0)=(0.1,1.0)$. In all cases, the largest adsorption takes place at the bottleneck ($\alpha$). For high adsorption rates and low diffusion, the cage ($\beta$) adsorbs at higher rate than the throat ($\epsilon$). For low adsorption rates and fast diffusion, both the cage and the throat ($\epsilon$) adsorb at the same rate.

\paragraph{Diffusion, adsorption and chemical production in a pore with discrete distribution of active sites.}
The most general case that may be analyzed by the Fick-Jacobs generalized equation, Eq. (\ref{eq:fjparac}), corresponds to the case when diffusion, adsorption and chemical production take place in the bulk of the pore. 

\begin{figure}[]
\centering
\includegraphics[width=8.5cm]{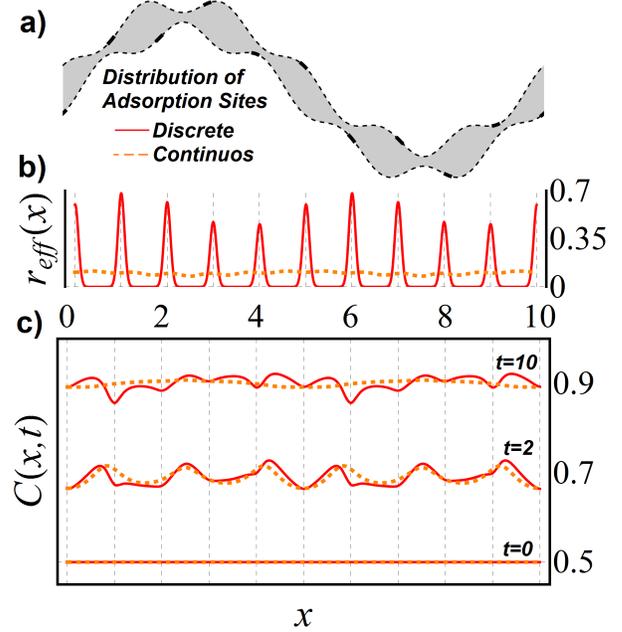}
\caption{ Ilustration of the predicted bulk concentration profiles of Eq. (\ref{eq:fjparac}) for a process including diffusion, adsorption and chemical reaction.  a) Sinusoidal shaped pore with $(L,\tau,\kappa)=(10,1.5,0.2)$ in Eqs. (\ref{eq:porefunctions}). We consider two situations: i) the active sites of adsorption are distributed in discrete sites remarked with thick solid lines, and ii) they are distributed homogeneously over the two walls (dashed lines). b) Effective rate of adsorption for these two cases are $r_{eff}=0.5\,C(x,t) \gamma(x)$ and $r_{eff}=0.5\,C(x,t)  \sum_{k=1}^{10}\exp[-(x-k)^2/\sigma^2]\gamma(x)$, respectively. c) Concentration profiles predicted by Eq. (\ref{eq:fjparac}), with $D_0=0.5$, and $k_f=0.5$ and $g$ given by Eq. (\ref{ultima}). Vertical dashed lines represent the location of active sites for the discrete case.\label{figure8}}
\end{figure}

The adsorption reaction at the pore's wall is described by  Eq. (\ref{eq:radsor}) with $r_{ads}$ given in two different ways. In the first one, we will assume a discrete distribution of active sites, remarked with thick solid lines over the pore in Fig.  \ref{figure8}a). In this case, the effective rate of adsorption  $r_{eff}(x)=\gamma(x)r_{ads}(x)$ is given as a sum of Gaussian functions, see Eq. (\ref{eq:kdesorpt2}), multiplied by the geometric factor $\gamma$. It is represented by the solid line in Fig. \ref{figure8}.b). In the second case, we have considered a continuous distribution of active sites over the same pore. The effective adsorption rate is represented by the dashed line in Fig. \ref{figure8}.b).  

The bulk reaction term $g$ was modeled (to take an example) by the kinetics
\begin{equation}\label{ultima}
g(x,t)=k_f C(x,t)[1-C(x,t)],
\end{equation}
where $k_f$ is the formation velocity constant of the species that will be adsorbed and $C(x,t)$ its bulk concentration.  We should remark that the generalized Fick-Jacobs equation (\ref{eq:fjparac}), with $g$ given by (\ref{ultima}) can give place to the formation and propagation of traveling waves in irregular confined geometries \cite{martens2015front}. However, in this work,  we will focus only in the effects of active sites of adsorption.

The resulting bulk concentration profiles for the two cases considered at three different times are shown in Fig. \ref{figure8}.c).  Solid lines are the predictions of the generalized Fick-Jacobs equation in the discrete case, whereas dashed orange lines correspond to the continuous case. As it can be seen on Fig. \ref{figure8}.c), the concentration profile can reveal information about the distribution of active sites, and also about the geometric disposition of the pore.

In order to provided a detailed description of the localization of active sites in an experimental setup, one has to be capable of obtain the full concentration profile as (function of $x$ and $y$).  In the following section, is discussed how the model presented in this work could serve to improve the obtention of this two-dimensional concentration profiles, at least, for the equilibrium case.

\section{Conclusions and Perspectives\label{sec:conclusion}}

In the present work, we have explored the range of applicability of the generalized Fick-Jacobs equation in the case when diffusive mass transport of a fluid along a pore includes chemical reactions in the bulk and pore's surface. The obtained equation permit us to calculate the concentration profiles for many situations involving these processes, and shows excellent agreement with the predictions obtained by using the two dimensional mass balance equation when the length of the pore is much longer than the width, even for non-equilibrium boundary conditions. One of the merits of the present study is that it establishes the exact mathematical relations between these two descriptions and, as a consequence of this, it establishes a theoretical bridge that connects the measurements of the average quantities with the more detailed processes taking place at the surface of the pore.

Our geometric derivation of the generalized Fick-Jacobs equation Eq. (\ref{eq:fjparac}) permits one to distinguish the relevant geometric aspects of the diffusion and the adsorption for structured porous media, fact that would permit us to find {\it a priori} the better design for an engineered material. This important contribution goes beside of the computational easiness of solving the one-dimensional projected equation when compared with the numerical solution of the two dimensional one. In practice, our model can be used as a predicting tool in chemical reactor problems where the internal diffusion inside the pore media or the adsorption-desorption kinetics are the limiting processes. Our numerical calculations permit us to corroborate these conclusions both  in equilibrium as well as in non-equilibrium situations, for both continuous and discrete distributions of active sites.

From our analysis, it follows that the tortuousness and constriction of the pore can be well represented by a reduced scheme. In the case of the processes of adsorption and desorption, we have proven (analytically and numerically) that the effective length of the pore walls is the geometric relevant quantity in this type of reactors. Hence, we corroborated that the local curvature of the cages of a pore as well as its constriction can be very important factors in the manufacture of fluid sieves.

Besides the theoretical aspects, we believe that the major contribution of this work is the fact that it would permit us to establish some new basis of comparison between a mean field model, as the one of us, with works related with the chemical engineering of reactors, since the proposed scheme can provide relations with quantities such as effective  rates of adsorption and  effective diffusion coefficients  \cite{ carberry1987chemical,fogler1999elements,cussler2009diffusion}. In this context, as we have shown with the help of very simplified examples, the model could serve to understand the reactor yield in a new diversity of structured porous media,  where the average shape of the empty space is well represented by long tube-like structures. Besides, we believe that this work can be extended to an interconnected net of tubes as well as to granular media using some cylinder-cell model \cite{coutelieris2012transport}.

 A very interesting idea to explore in the future is the fact that the effective efficiency of adsorption in a given active site depends specifically on the ratio $ r_i /A_i$,  where $A_i(x,t)$ and $r_i(x,t)$ are the fluxes in Eq. (\ref{eq:Jtrabajado2}), i.e. between the rate of the diffusion flow and the rate of adsorption near the adsorption site. In order to determine this efficiency, it is necessary to know explicitly the complete two dimensional concentration profile $C(x,y,t)$. However, since the projected schemes that we use does not contain specific information about this two-dimensional concentration profile, a study in the non-equilibrium regime has to be performed in order to  know the specific values of the concentration near the walls of the pore and, therefore, to establish the adsorption efficiency inside an irregular pore with more accuracy. In this sense, our model provides a first approximation in terms of averaged quantities . A second approximation to the problem would consist in obtaining the entire concentration profile along similar lines to those of  Refs. \cite{kalinay2005projection,martens2011entropic,pineda2014projection}. In principle, this would permit to estimate the value of the concentration near the wall and, therefore, the adsorption/desorption rates with more precision. However, these approximations only provide the equilibrium concentration profile. Because we are interested in non-equilibrium situations, we consider that it is more consistent the use of the averaged time-dependent concentration profile than the equilibrium one. This is why the use of averaged quantities (as explained in Section \ref{sec:average}) is crucial for the correct interpretation of the obtained results.

Another perspective of future work is related to the diffusion coefficient used. The obtention of a modified effective diffusion coefficient that includes the presence of entropic barriers as well as the effects of chemical reactions is a necessary step in order to include more general kinetic mechanisms than those considered in this work. On relation with this, at least three strategies can be adopted that may use the equilibrium concentration profile determined by equation (27): $a$) as a zero order solution of an asymptotic projection scheme\cite{kalinay2005projection,martens2011entropic}; $b$) as the base of a fictitious initial condition in the context of a macro-transport theory\cite{shapiro1986taylor,shapiro1988dispersion}; and $c$) using the stationary or equilibrium concentration profiles predicted by the chemical kinetics within a self-consistent non-equilibrium thermodynamics scheme\cite{santamaria2012non,santamaria2013entropic}.


%
%


\bibliographystyle{abbrv}
\bibliography{chap2v3.bib}

\end{document}